\documentclass[a4paper]{jpconf}
\usepackage{graphicx}
\usepackage{psfrag}
\usepackage{epsfig}
\usepackage{amsmath}
 \usepackage{amscd}
 \usepackage{hyperref}
 \usepackage{amsthm}
 \usepackage{amssymb}
\usepackage{epsf,float}
\usepackage{bm}%

\begin{document}
\title{Intrinsic periodicity: \\ the forgotten lesson of quantum mechanics\footnote{Dedicated to Ignaz Philipp Semmelweis, the ``savior of mothers'' (Buda 1818 --- D\"obling 1865).}}

\author{Donatello Dolce}

\address{  
  the University of Melbourne,   Parkville 3010 VIC, Australia. 
\\ 
the University of Camerino, Piazza Cavour 19F, 62032 Camerino, Italy.}

\ead{donatello.dolce@coepp.org.au}

\begin{abstract}
Wave-particle duality, together with the concept of
elementary particles, was introduced by de Broglie in terms of intrinsically Òperiodic phenomenaÓ. However, after nearly 90 years,
the physical origin of such undulatory mechanics remains unrevealed. We propose a natural realization of the de Broglie Òperiodic phenomenonÓ in terms
of harmonic vibrational modes associated to space-time periodicities. In this way we find that,
similarly to a vibrating string or a particle in a box, the intrinsic recurrence imposed as a constraint to elementary particles represents
a fully consistent quantization condition. The resulting classical cyclic dynamics formally 
match ordinary relativistic Quantum Mechanics in both the canonical and Feynman formulations.
Interactions are introduced in a geometrodynamical way, similarly to general relativity, by simply considering that variations of kinematical state can be equivalently described in terms of modulations of space-time recurrences, as known from undulatory mechanics. We present this novel quantization prescription from an historical prospective.    
\end{abstract}

\section{Introduction}

We give some historical motivation for a formulation of Quantum Mechanics (QM) in terms of elementary space-time cycles \cite{Dolce:2009ce,Dolce:tune,Dolce:AdSCFT}. Since its earliest days, QM was characterized by an assumption intrinsic periodicity. We may think to the  wave-particle duality, introduced by de Broglie in this terms: ``we proceed with the assumption of the existence of a certain periodic phenomenon of a yet to be determined character, which is to be attributed to each and every isolated energy parcel'' \cite{Broglie:1924}. That is to say, every elementary particle is characterized by an intrinsic recurrence in time and space. As known from undulatory mechanics, every local retarded variation of the kinematical state of a particles (i.e. the local four-momentum) can be equivalently expressed, by means of the Planck constant, in terms of the corresponding local retarded modulation of space-time recurrence.  According to the Standard Model, every system in physics is described by a set of elementary particles and their relativistic interactions, thus the wave-particle duality implicitly says that every system in physics can be described in terms of elementary modulated cycles.
 
The intrinsic periodicity of elementary particles can be used as a quantization condition, similarly to the walls of a ``particle in a box''. This is intuitive if we think to Bohr's hydrogen atom in which, similarly to the harmonic spectrum of a vibrating `string', the energy levels are determined by requiring that the only possible orbits are those with an integer number of wavelengths. Indeed this is the idea at the base of the Bohr-Sommerfeld quantization: in a semi-classical description of QM, the energy spectrum can be determined by requiring closed orbits associated to a given potential. 
It has been shown recently that such a semi-classical description can be used to solve complex quantum systems such as atoms with more electrons or the Zeeman effect, considered as some of the main limitations of this approach.   

More generally, as mathematically proven in recent papers \cite{Dolce:2009ce,Dolce:tune,Dolce:AdSCFT}, we will show how the semi-classical description based on the assumption of intrinsic periodicity of elementary systems, when correctly implemented in a relativistic theory, can be used to formally derive  the modern descriptions of relativistic QM.

\section{The model}

We introduce the formalism by following the same arguments used by de Broglie to derive the wave-particle duality \cite{Broglie:1924}. In his paper he pointed out that every elementary particle is characterized by recurrences in time and space. These are fully determined, though Lorentz transformations, by the recurrence along the proper time, fixed by the mass of the particle $\bar M$.  The periodicity of this rest recurrence   is the Compton time of the particle 
$$T_{\tau} = \frac{h}{ \bar M c^{2}}.$$ 
Equivalently, every particle has a quantum recurrence $\lambda_{s}$ of along its worldline $s= c \tau$, where $\lambda_{s}$ is the Compton wavelength $\lambda_{s} = c T_{\tau} $. When observed in a generic reference frame this corresponds to instantaneous temporal and spatial periodicities,  $T_{t}$ and $\vec \lambda_{x}$ respectively, as can be seen by using the Lorentz transformation 
$$c T_\tau= c \gamma T_{t} - \gamma \vec \beta \cdot \vec \lambda_{x}.$$ 
They can be written in a contravariant tangent four-vector $$T^{\mu}=\{T_{t},\vec \lambda_{x}\}.$$ In fact, from the relativistic definition of four-momentum $\bar p_{\mu}=\{\bar E ,-\mathbf{\bar{p}}\}= \{\gamma \bar M c^{2} ,-\gamma \vec {\beta} \bar M c\}$, it is easy to derive the following relativistic invariant (de Broglie phase harmony)  
$$T_\tau \bar M c^2 \equiv h ~~~\longleftrightarrow~~~ T^{\mu} {\bar{p}}_{\mu}  \equiv h. $$  
In other words, the four-momentum of a particle can be equivalently encoded in corresponding modulations of the instantaneous space-time periodicity $T_{\mu}$. According to undulatory mechanics, particles can be described  in terms of periodic phenomena (phasors or waves) in which the space-time coordinates enter as angular variables; their periodicities define their kinematical state. Thus, every system in nature can be consistently described in terms of modulations of elementary space-time cycles. Since they are completely characterized by proper time periodicity, we can consider a minimal topologies $\mathbb S^1$ for  elementary bosonic cycles\footnote{We neglect the spin which however can be regarded as another manifestation of the intrinsic periodicity of elementary particles. See Schr\"odinger's \emph{zitterbewegung} and related semiclassical derivation of the Dirac equation.}. 

It must be noted that these recurrences are typically extremely fast with respect to the Cs$_{133}$ atomic clock, whose time scale is (by definition of ``second'') about $10^{-10} s$, or to the present resolution in time, about $10^{-17} s$ (using laser pulses). For instance, simple electrodynamics system are characterized by the the proper time periodicity of the electron $T_{\tau (e)} = h / m_{(e)} c^{2} = 8.09329972 \times 10^{-21} s$ (Compton time), corresponding to the \textit{zitterbewegung}. The heavier the mass or the energy of the particles, the faster the corresponding temporal periodicities. For instance, LHC is exploring ``indirectly''\footnote{With ``direct'' observation we mean to compare these fast dynamics with a clock of resolution higher than this time scale. For instance, with the modern clocks, a direct observation of these quantum recurrences are possible for neutrinos and other coherent quantum phenomena in condensed matter such as lasers, superconductivity or carbon nanotubes.} time cycles of the order of $T_{t} \sim 10^{{-27}} s$ (corresponding to energies scales of the TeV).

We want to impose the intrinsic periodicity $T^{\mu}$ of elementary particles  as a constraint. This represents a semiclassical quantization condition. A particle with intrinsic periodicity is similar to a ``particle in a box''. Through discrete Fourier transform the periodicity $T^{\mu}$ directly implies a quantization of the conjugate spectrum $p_{n}^{\mu} = n \bar p^{\mu}$; $n$ is the single quantum number associated to the topology $\mathbb S^{1}$. For instance the quantization of the energy spectrum associated to the time periodicity $T_{t}$ is the harmonic spectrum 
$$E_{n} = n \bar E = n h / T_{t}.$$ 
A bosonic particle can be therefore represented as a one dimensional bosonic `string' $\Phi(x)$ vibrating  in compact space-time dimensions of length $T^{\mu}$ and Periodicity Boundary Conditions (PBCs --- denoted by the circle in $\oint$):
\begin{equation}
  {\mathcal{S}}^{\lambda_{s}} = \oint^{T^\mu}  d^4 x {\mathcal{L}}(\partial_\mu \Phi(x),\Phi(x)) \,.\label{generic:actin:comp4D}
\end{equation}
As known from string theory or extra-dimensional theories, PBCs (or  combinations of Dirichlet/Neumann  BCs or anti-PBCs) minimizes the action at the boundary so that all the relativistic properties of (\ref{generic:actin:comp4D}) are preserved. This is a consequence of the fact that relativity fixes the differential structure of space-time whereas the only requirement for the BCs is to fulfill the variational principle.  The expansion in harmonics (discrete Fourier expansion with coefficients $a_{n}$ and normalization $A_{n}$) of a field/string vibrating with persistent periodicity is 
$$\Phi(x) = \sum_{n} \phi_n(x) = \sum_{n} A_{n} a_{n}(\bar{p}_{\mu}) e^{{-\frac{i}{\hbar} p_{n \mu} x^{\mu}}}. $$  

Such a description in terms of elementary space-time cycles is covariant as prescribed by relativity. This can be easily seen by using Lorentz transformations. The action (\ref{generic:actin:comp4D}) describing a particle in a given reference frame, after the generic transformation of variables  $x_\mu \rightarrow x'_\mu = \Lambda_\mu^\nu x_\nu$,  transforms to the action
\begin{equation}
  {\mathcal{S}}^{\lambda_{s}}  = \oint^{T'^{\mu} =\Lambda^\mu_\nu T^\nu} d^4 x' {\mathcal{L}}( \partial'_\mu \Phi'(x'),\Phi'(x'))\,. \label{lorentz:actin:comp4D}
\end{equation}
This turns out to have a transformed boundary, such that the resulting solution has transformed periodicity $$T^\mu \rightarrow {T'}^\mu = \Lambda^\mu_\nu T^\nu.$$ According to $\bar p'_{\mu} c T'^{\mu} = h$, this is the periodicity associated to the four-momentum  of the particle in the new frame
$$\bar p_\mu \rightarrow  {{\bar p}'}_\mu = \Lambda_\mu^\nu \bar p_\nu \,.$$
 Indeed $T^{\mu}$ is a contravariant tangent four-vector satisfying the relativistic constraint  
 $$\frac{1}{T^{2}_{\tau}} = \frac{1}{T_{\mu}} \frac{1}{T^{\mu}},$$ 
which is the geometric counterpart of the relativistic dispersion relation of the particle $\bar M c^{2} = \bar p_{\mu} \bar p^{\mu}$. By considering  the relativistic modulations of temporal periodicity (relativistic Doppler effect)   associated to variations of reference frames, the resulting energy spectrum is 
$$E_{n} (\bar {\mathbf{p}}) = n \frac{h}{ T_{t} (\bar{ \mathbf{p}})} = n \sqrt{\bar{ \mathbf{p}}^{2} c^{2} + \bar M^{2} c^{4}}.$$
 In other words we have obtained semi-classically the energy spectrum prescribed by ordinary second quantization (after normal ordering) for bosonic particles\footnote{In fact, second quantization prescribes that every mode with angular frequency  $\bar \omega (\bar{ \mathbf{p}}) = \sqrt{\bar{ \mathbf{p}}^{2} c^{2} + \bar M^{2} c^{4}} / \hbar$  of a Klein Gordon fields has a quantized energy spectrum $E_{n} (\bar {\mathbf{p}}) = n h \bar \omega (\bar{ \mathbf{p}})$, after normal ordering. By equivalently assuming in (\ref{generic:actin:comp4D}) anti-PBCs we get $E_{n} (\bar {\mathbf{p}}) = (n + \frac{1}{2}) h \bar \omega (\bar{ \mathbf{p}})$, avoiding normal ordering.}. 
 

\section{Interactions}

Interactions can be introduced in this formalism by considering that, as known in undulatory mechanics, relativistic variations of four-momentum associated to a given interaction scheme can be equivalently described in terms of relativistic modulations of periodicities. That is, in every point $x=X$, a relativistic interaction can be characterized by the relativistic variations of four-momenta of a particle with respect to the free case 
$$\bar{p}_{\mu}\rightarrow\bar{p}'_{\mu}(X)=e_{\mu}^{a}(x)|_{x=X}\bar{p}_{a}.$$ 
Through $\hbar$, this generic interaction can be equivalently encoded by corresponding relativistic modulations of quantum recurrence 
$$T^{\mu}\rightarrow T'^{\mu}(X)\sim e_{a}^{\mu}(x)|_{x=X}T^{a}.$$
 According to our formalism, this corresponds to local and retarded ``stretching''  the compactified space-time dimensions of (\ref{generic:actin:comp4D}), and thus by corresponding local and retarded deformations of the metric $$\eta_{\mu\nu}\rightarrow g_{\mu\nu}(X)=[e_{\mu}^{a}(x)e_{\nu}^{b}(x)]|_{x=X}\eta_{ab}.$$ 
 This geometrodynamical description of interaction can be easily checked by imposing the local transformation of reference frame 
 $$dx_{\mu}\rightarrow dx'_{\mu}(X)=e_{\mu}^{a}(x)|_{x=X}dx_{a}$$ 
 as substitution of variables in the free action (\ref{generic:actin:comp4D}), see \cite{Dolce:tune,Dolce:AdSCFT}. Indeed, the resulting action turns out to have locally deformed metric $g_{\mu\nu}(X)$ and local transformations of boundary. In this way, the solution of this transformed action correctly describes the local modulations of  periodicity $T'^{\mu}(X)$ associated to the interaction scheme. That is, we pass from a free solution of persistent type  
 $$\phi(x) \propto e^{-\frac{i}{\hbar} p_{ \mu} x^{\mu}} $$  
 to the interacting solution with modulated periodicity of type  
 \begin{equation}
 \phi'(x) \propto e^{-\frac{i}{\hbar} \int^{x_{\mu}} d x'^{\mu} p_{\mu} }. \label{inter:mode}
 \end{equation} 
 Note also that in our formalism the (integral of the) kinematics associated to the given interaction scheme turns out to be encoded on the boundary of the theory. The holographic principle has thus a natural realization in this theory. 
 
Such a geometrodynamical description of generic interactions is of the same type of the one prescribed by General Relativity (GR).  In a weak Newtonian interaction, the corresponding variation energy  $\bar{E}\rightarrow\bar{E}'\sim\left(1+{GM_{\odot}}/{|\mathbf{x}|c^2}\right)\bar{E}$  implies, through $\hbar$, a modulation of time periodicity $T_{t}\rightarrow T_{t}'\sim\left(1-{GM_{\odot}}/{|\mathbf{x}|c^2}\right)T_{t}$, i.e. we have redshift and time dilatation. If we also consider the variation of momentum and the corresponding modulation of spatial periodicity, the resulting metric encoding the Newtonian interaction is actually the linearized Schwarzschild metric.  Thus we have obtained linearized gravity, from which GR follows by assuming self-interaction. 

We also mention another remarkable consequence of our description of elementary cycles, which can be regarded as a realization of Weyl's, Kaluza's, Wheeler's original proposals. In  \cite{Dolce:tune,Dolce:AdSCFT} we have found that gauge interaction has a geometrodynamical description similar to that of gravitational interaction in GR. This possibility is provided by the fact that transformations of flat reference frames, which in ordinary QFT do not vary the solution of the theory\footnote{In ordinary QFT the BCs have a marginal role:  the KG field used in computation is the most general solution of the KG equation.}, in our formulation induce local transformations of the boundary, and thus local transformations of the solution of the theory. The resulting transformation of the field solution is formally equivalent to the internal transformation of ordinary gauge theories. For instance, it is possible to show that the local transformation of flat reference frame\footnote{Intuitively we want to describe the trembling motion of a particle interacting electromagnetically as local transformation of reference frame, similarly to the equivalence principle (applied to the boundary) of GR. } $$dx^{\mu}(x)\rightarrow dx'^{\mu}\sim dx^{\mu} - e dx^{a} \omega^{\;\mu}_{a}(x)$$ can be parametrized parametrizing by introducing a vectorial field $$\bar{A}_{\mu}(x) \equiv \omega_{\;\mu}^{a}(x)\bar{p}_{a},$$ so that the interaction scheme is actually given by the minimal substitution $$\bar{p}'_{\mu}(x)  \sim  \bar{p}_{\mu}- e\bar{A}_{\mu}(x).$$ Thus the transformed solution associated to this interaction is of the type $$\phi'(x) \propto e^{-\frac{i}{\hbar} \int^{x_{\mu}} d x'^{\mu} A_{ \mu} } e^{-\frac{i}{\hbar} p_{ \mu} x^{\mu}}. $$ The gauge connection, which must be postulated in ordinary gauge theory, turns out to describe the modulations of periodicity associated to the local transformations of flat reference frame with respect to the free solution (we say that the gauge field ``tunes'' the periodicity). From this, ordinary Maxwell dynamics (and more in general Yang-Mills theories) can be derived,  see \cite{Dolce:tune} for more details.

\section{Quantization}

From an historical point of view, the modern mathematical formulation of QM is inspired to the theory of sound and Reyleigh studies. Indeed the theory described so far can be regarded as a fully relativistic generalization of the theory of sound (we assume vibrations in compact time, and not only in compact space as in ordinary classical sound sources). Therefore the theory inherits fundamental aspects of the ordinary quantum formalism. The assumption of periodicity at the base of our formulation is a quantization condition (similarly to the quantization of a particle in a box). In particular, when the recurrences of elementary particles are imposed as a constraint, the resulting (classical) cyclic dynamics reproduce formally ordinary relativistic QM. 

In this section we summarize basic aspects of this remarkable correspondence. It is known that a vibrating `string'  is the typical classical system that can be described locally in a Hilbert space. Even if we consider modulations of periodicities, the harmonics of such an interacting `string'  form locally a complete set with respect to  the corresponding local inner product \begin{equation}
\left\langle \phi|\chi\right\rangle \equiv\int_{0}^{\lambda_{x}(X)}\frac{{dx}}{{\lambda_{x}(X)}}\phi^{*}(x)\chi(x)\,.\label{inner:prod}\end{equation}
These harmonics define locally a Hilbert base $$\left\langle x | \phi'_{n}\right\rangle = \phi'_{n}(x).$$ Thus a modulated vibrating `string', generic superposition of harmonics, is represented  by the generic Hilbert state $$ \left| \phi' \right\rangle = \sum a_{n} \left| \phi_{n} \right\rangle.$$ The non-homogeneous Hamiltonian $\mathcal H'$ and momentum $\mathcal P'_{i}$ operator are introduced as the operators associated to the four-momentum spectrum of the locally modulate `string': $$\mathcal P'_{\mu}\left |\phi'_{n} \right\rangle = p'_{n \mu}  \left|\phi'_{n} \right\rangle, ~~\text{where} ~~ \mathcal{P}'_\mu = \{\mathcal H', - \mathcal{P}'_{i}\}\,.$$  From the modulated wave equation, the temporal and spatial evolution of every modulated harmonics satisfies $$i \hbar \partial_{\mu} \phi'_{n}(x) = p'_{n} \phi'_{n}(x),$$ thus the time evolution of our modulated `string' $ \left| \phi' \right\rangle $  is given by the ordinary Schr\"odinger equation 
\begin{equation}
i \hbar \partial_{t} \left|\phi' \right\rangle = \mathcal H' \left|\phi \right\rangle. \label{schro:eq}
\end{equation} 
Moreover, since we are assuming intrinsic periodicity, this classical-relativistic theory implicitly contains the ordinary commutation relations of QM. This can be seen by evaluating the expectation value of a total derivative $\partial_{x} F(x)$, and considering that the boundary terms of the integration by parts cancel each other owing the assumption of intrinsic periodicity.  For generic Hilbert states we \emph{obtain} $ [F(x),\mathcal{P}'_{i}] = i \hbar \partial_{x} F(x) $ and thus, from $F(x)=x_{j}$, $$ [x_{j},\mathcal{P}'_{i}] =i \hbar \delta_{j,i} \,.$$
 The formal correspondence to ordinary relativistic QM is confirmed by the fact that, remarkably, the classical evolution of such a classical vibrating `string' with all its modulated harmonics is described by the ordinary Feynman Path Integral (we are integrating over a sufficiently large number $N$ of spatial periods so that the $V_{\mathrm{x}} = N \lambda_{x}$ is bigger than the interaction region)
 \begin{equation}
\mathcal{Z}=\int_{V_{\mathrm{x}}} {\mathcal{D}\mathrm{x}}  e^{\frac{i}{\hbar}  \mathcal{S}'(t_{f},t_{i})}\,.\label{eq:Feynman:Path:Integral}\end{equation}
The action $\mathcal S'$ is, by construction, the classical  action of the corresponding interaction scheme, with lagrangian $\mathcal{L}' = \mathcal P' x - \mathcal H' $. This result has a very intuitive justification in the fact that in a cyclic geometry such as that associated to the topology $\mathbb S^{1}$, the classical evolution of $\phi(x)$ from an initial configuration to a final configuration  is given by the interference of all the possible classical paths with different windings numbers; without relaxing the classical variational principle.  Thus the harmonics of the vibrating string/field $\phi(x)$ are interpreted as quantum excitations. We have also proven that this classical to quantum correspondence pinpoints a fundamental aspect of Maldacena's duality \cite{Dolce:tune}. Therefore, the assumption of intrinsic periodicity which has characterized quantum theory in its early days has a renewed interest in modern physics.
This correspondence to relativistic QM can be interpreted in analogy with 't Hooft determinism: ``there is a deep relationship between a particle moving very fast in a circle [of periodicity $T_{t}$] and a quantum harmonic oscillator [with the same periodicity]'' \cite{'tHooft:2001ar}; or the stroboscopic quantization \cite{Elze:2003ws}). QM emerges as a statistical description associated to the extremely fast recurrences of the elementary systems in nature: this is the meaning of the Hilbert notation above.  Our actual experimental time resolution, about $\Delta T_{exp}  \sim 10^{-17}$,  is too low to resolve the small time scales of ordinary quantum systems, though the internal clock of an electron has been observed in a recent interference experiment \cite{2008FoPh...38..659C}.   As for a dice rolling too fast with respect to our resolution in time, the outcomes can be described only in a statistical way. Loosely speaking, an observer with infinite resolution in time can in principle resolve exactly the underling deterministic cyclic dynamics: it would have no fun playing dice (``God doesn't play dice'', Einstein).

It is instructive to interpret the black-body radiation in terms of ``periodic phenomena''. In this case it is natural to assume that the population $a_n$ of the $n^{th}$ energy levels is fixed by the Boltzmann  distribution. For the IR components of the radiation, i.e. massless periodic phenomena with long periodicity, the PBCs can be neglected and the energy spectrum can be approximated to a continuum (classical limit) since many energy levels are populated (the thermal noise, i.e. the euclidean time periodicity, destroys the intrinsic periodicity in a sort of decoherence). For  the UV components, however,  the fundamental energy $\bar E$ is big with respect to the thermal energy so that only a few energy levels can be populated. That is to say, these modes have very short periodicity (not destroyed by the thermal noise), so that the PBCs are important and there is a manifest quantization of the energy spectrum (quantum limit). The UV catastrophe is avoided according to Planck.    

We have a consistent interpretation of quantum to classical transition as $\hbar \rightarrow 0$.
For massive particles, in the non-relativistic limit $\bar p \ll \bar M c$, only the fundamental energy level is largely populated as the gap between the energy levels goes to infinity, $\bar M \rightarrow \infty$. 
The cyclic field can therefore be approximated as  $ \Phi(x) \sim \exp[-i \frac{\bar M c^{2}}{\hbar} t + i \frac{\bar M}{\hbar} \frac{\mathbf x^{2}}{2 t}]$, see \cite{Dolce:2009ce}. Neglecting the proper tim recurrence, it is possible to see by plotting the $\left| \right|^{2}$ that the wavefunction of a massive periodic phenomenon is centered along the path of the corresponding classical particle and its width is of the order of the Compton wavelength $\lambda_s$. Thus, in the non-relativistic limit, the Dirac delta distribution describing a classical particle is reproduced  as in the usual Feynman description. Similarly, the spatial compactification length tends to infinity whereas the time compactification tends to zero, so that in the non-relativistic limit we have a point-like distribution in $\mathbb R^3$, i.e. the ordinary three-dimensional description of a classical particle. Indeed, the corpuscular description therefore arises at high frequencies.        
On the other hand, in the relativistic limit the non-local nature of a massive periodic phenomena can not be neglected (the distribution width is  of the order of the Compton wavelength). This gives an intuitive interpretation of the wave-particle duality and of the double slit experiment. In particular, if probed with high energy or observed with good resolution, more and more energy levels  turn out to be excited, i.e. more and more harmonics can be resolved. In this way we can figure out that the energy excitations play the role of the quantum excitations of the same fundamental elementary system, so that we have an analogy with the \emph{virtual} particles of ordinary relativistic QM.  The modes can have in general with either positive or negative frequencies: this means that the relativistic theory has Hamiltonia  operator positively defined (contrarily to 't Hooft model). 

We conclude this section by noting that the assumption of intrinsic periodicity implicitly contains
the Heisenberg uncertain relation of ordinary QM. Briefly, in the Hilbert notation (statistical description of a ``periodic phenomenon'')  the phase of the periodicity can not be determined --- in (\ref{inner:prod}) only the square of the field has physical meaning . To determine the frequency and thus the energy
$\bar{E}(\mathbf{\bar{p}})=\hbar \bar{\omega}(\mathbf{\bar{p}})$ with good accuracy $\Delta\bar{E}(\mathbf{\bar{p}})$ we
must count a large number of cycles. That is to say we must observe the system for a long time $\Delta t(\mathbf{\bar{p}})$.    The phase is defined modulo factors $\pi$ which can be written as a energy or a temporal uncertainty $\exp[-i\frac{\bar E t}{\hbar} + \pi] = \exp[-i\frac{(\bar E + \Delta\bar{E})  t}{\hbar} ] = \exp[-i\frac{\bar E  ( t + \Delta t)}{\hbar} ]$. Considering that every cycles is such that $t < T_{t}$ (for time longer than a cycle we must consider the general phase invariance $n \pi$)  we find that this uncertain is described by the famous Heisenberg relation $\Delta\bar{E}(\mathbf{\bar{p}})\Delta t(\mathbf{\bar{p}})\gtrsim\hbar/2$,   \cite{Dolce:2009ce}.
  
\section{Schr\"odinger problems}

We continue with the historical overview of our assumption of intrinsic periodicity --- aspects concerning modern physics are discussed in \cite{Dolce:2009ce,Dolce:tune,Dolce:AdSCFT}. In particular we what to show how to solve simple non-relativistic quantum problems in terns of the vibrational modes associated to the quantum recurrences\footnote{More in general, by quantum recurrence we mean that quantum system can be described semi-classically by imposing  appropriate BCs to relativistic wave equations. Trivial examples are the quantization of a particle in a box, a particle with Dirac delta potentials, and similar problems. However this approach can be extended to non-obvious problems such as the Casimir effect \cite{Jaffe:2005vp}, the tunnel effect or atomic physics \cite{Solovev:2011}}. 

First of all we must consider that the assumption of intrinsic periodicity is equivalent to the Bohr-Sommerfeld condition: it in fact prescribes that the only possible modes are those with a finite number of cycles as in a vibrating `string'. For instance, the assumption of intrinsic periodicity in the free case (homogeneous periodicity) leads to the harmonic spectrum $p_{n \mu} T^{\mu}= n \bar p_{ \mu} T^{\mu} = n h$. More in general, as shown in \cite{Dolce:tune}, in case of interaction the quantized spectrum associated with the corresponding modulation of periodicity is given by the condition $\oint p_{n \mu}(x) d x^{\mu} = h (n + v) $ (the so-called Morse index $v$ interpreted as a twist factor in the PBCs, see also \cite{Dolce:2009ce,Dolce:tune,Solovev:2011}). This can be seen by imposing periodicity to the modulated solution of the type (\ref{inter:mode}); note also the correspondences with the WKB method. To obtain the the Bohr-Sommerfeld we quantized only the spatial momentum $\mathbf{p}_{n}(\mathbf{x}) $; the corresponding energy spectrum is obtained through the equations of motion of the interaction. The quantization condition is therefore $$\oint \mathbf{p}_{n}(\mathbf{x}) d \mathbf{x} = h (n + v) .$$ To this spatial momentum we associate a potential, which in the non-relativistic limit can be defined such that $ E_{n} = \frac{\mathbf{p}^{2}_{n}(\mathbf{x})}{2 m} +  V(\mathbf{x}) $. The Hamiltonian operator is so that the Schr\"odinger equation (\ref{schro:eq}), mode by mode, turns out to be given by 
\begin{equation}\label{harm:oscill:eigen}
- \frac{\hbar^2}{2 m} \partial_\mathbf{x}^2 \phi_n(\mathbf{x})+ V(\mathbf{x}) \phi_n(\mathbf{x}) = E_n \phi_n(\mathbf{x}) ~.
\end{equation}

Taking into account all  previous arguments,  the quantized harmonic oscillator turns out to have an immediate solution\footnote{This avoides to the complication of the Fock space.  In this approach the creation and annihilation operators  can be regarded as describing the excitation or de-excitation of the different levels of the harmonic system.} with respect to the usual formulation. It is sufficient to consider the (Galileo) isochronism of the pendulum: every orbit of an harmonic oscillator has the same time (homogeneous) periodicity. We have just to impose that the solution is a wave with the characteristic periodicity  $T_{t} = 2 \pi / \bar \omega $ for every energy level. That is the quantization condition of the energy spectrum is simply $\oint E_{n \mu} d t = E_{n} T_{t} = h (n + v)$. Therefore the solution has an harmonic energy spectrum $E_{n} = \frac{h}{T_{t}} (n + \frac{1}{2})$ and  can  be decomposed as 
\begin{equation}\label{HO:sol:gen}
\Phi(\mathbf{x},t) = \sum_n \phi_n(\mathbf{x},t) =  \sum_n e^{i (n + 1/2) \bar \omega t} \phi_n(\mathbf{x})  ~.
\end{equation}
where we have assumed a twist of half a period (anti-periodicity) in the BCs in order to reproduce the Morse factor $v= \frac{1}{2}$.\footnote{Intuitively the assumption of intrinsic antiperiodicity is natural for fermion in order to reproduce the correct spin statistics}
The Schr\"odinger differential equations of the problem can be written by defining  $\mathbf{x} = \rho \mathbf{y}$ with $\rho = \sqrt{\frac{\hbar}{m \bar \omega}}$ and $h_n(\mathbf{y}) = \phi_n(\mathbf{y}) e^\frac{ \mathbf{y}^2}{2}$. Thus the solution of the $n$-th harmonic mode of the vibration associated to the harmonic potential is
\begin{equation}
\partial_\mathbf{y}^2 h_n(\mathbf{y}) - 2 \mathbf{y} \partial_\mathbf{y} h_n(\mathbf{y}) = \left [2 \left(n + \frac{1}{2}\right ) - 1 \right ] h_n(\mathbf{y}) ~.
\end{equation}
We have solved the quantum harmonic oscillator by simply assuming intrinsic periodicity. This example is extremely important since the harmonic oscillator is the basic ingredient of ordinary (bosonic) quantum field theory. That is, a second quantized Klein Gordon field is the integral over the quantum harmonic oscillators with all the possible periodicities, from zero to infinity. The creation and annihilation operators in this approach describe the excitation of the different $n$-th mode of this harmonic `string'. This suggests that our correspondence to QM can be fully extended to ordinary quantum field theory.

In a similar way it is easy to solve other Schr\"odinger problems. For instance here we report the energy spectrum of simple  one-dimensional potentials. This method applied to the linear potential $V(x) = m g x$ actually gives the correct quantized energy spectrum $$E_n = \frac{1}{2} [3 \pi (n + 1/4) ]^{2/3} (\hbar^2 m g^2)^{1/3}$$ where we have assumed $v = 1/4$.  Another example is the solution of the quantum anharmonic oscillator, i.e. we assume a perturbation $\epsilon x^4/l^4$ of the harmonics potential, with $l = \sqrt{\hbar / m \omega}$. In this way we obtain (modulo Morse factors) the deviation for the harmonic energy spectrum $E_n = \hbar \bar \omega n $ of the same quantity prescribed by the ordinary methods: $$\Delta E_n = \epsilon \frac{3}{2}  (n^2 +  n).$$ In this way it is easy to see that the vibrational modes associated to these one-dimensional potentials reproduce the correct solutions of the Schr\"odinger problems.   

From a historical point of view this approach to quantum problems is the same adopted by Bohr to solve the energy spectrum of the hydrogen atom.  Indeed, according to the above prescription, the harmonic modes of our `string' with the modulated periodicities allowed by the Coulomb potential yields the resulting energy spectrum (modulo Morse factors)  $$E_{n} = -\frac{13.6 \text{eV}}{n^{2}}.$$ This famous results is obtained without necessarily requiring circular orbits. The single space-time periodicity of topology $\mathbb S^{1}$  associated to the orbits describes correctly the principal quantum number $n$. Furthermore, in spherical problems we must also consider the quantization associated to the spherical periodicity, i.e. the quantized vibrational modes associated to the additional periodicities in the spherical angles $\theta \in [0, 2\pi)$ and $\psi \in [0, \pi)$. As known from ordinary QM these lead to the ordinary quantization of  angular momentum. In fact, the spherical vibrational modes of solution  are described in terms of the  two additional angular and magnetic quantum numbers $\{m,l\}$. In other words, the atomic orbitals can be represented as the vibrational modes of an harmonic system of topology $\mathbb S^{1}\otimes \mathbb S^{2}$ in a Coulomb potential. This harmonic description of atoms has been recently re-proposed by eminent scientists such as the Nobel laureate Prof. F. Wilczek. It has been successfully generalized to achieve classical descriptions of quantum properties of atoms with more electrons (e.g. Helium atom), the Zeeman effect \cite{Solovev:2011}; these were considered the major limitation of the approach based on the assumption of intrinsic periodicity.

The example of the Hydrogen atom provides us the last element of the formal correspondence to QM: different harmonic sets characterized by different fundamental periodicity are described by the tensor products of their different Hilbert spaces. For instance, in the Hilbert notation,  the orbitals $\phi_{n,l,m}$, composition of the harmonics associated to the space-time vibration $\left| \phi_{n} \right\rangle$, or more simply $\left| n \right\rangle$, and the spherical harmonics  $\left| l,m \right\rangle$, can be written in a Hilbert space with base $\left| n \right\rangle \otimes \left| l,m \right\rangle = \left|n, l,m \right\rangle$ associated to the topology $\mathbb S^{1}\otimes \mathbb S^{2}$. 
This property plays an important role in the demonstration of the Bells' theorem. In particular, as it will be shown explicitly in a dedicated paper, the formal correspondence of our approach with canonical QM strongly suggests the statistical description of elementary space-time cycles satisfies the same inequalities of ordinary QM.   
In particular it must be noted that the theory has not local-hidden-variables; the assumption of intrinsic periodicity is a element of non-locality (though this type non-locality is consistent with relativity, as the periodicity varies according to the retarded potentials).

\section{Conclusions}

We have shown how ordinary relativistic QM can be formally derived from an effective statistical description of the classical mechanics associated to the space-time recurrences of elementary systems \cite{Dolce:2009ce,Dolce:tune,Dolce:AdSCFT}. In this paper we have given some historical motivations of this approach. It has its origin in Bohr's description of hydrogen atom, Bohr-Sommerfeld quantization, de Broglie's wave particle-duality, Schr\"odinger's \emph{zitterbewegung}, and in other foundational contributions to modern QM (not mentioned here for brevity). These ideas are the base of modern quantum field theory, but the attempt of a semi-classical description of QM has been abandoned after Heisenberg and Bell. Nevertheless, we have seen that these limitations can be avoided by thinking in terms of elementary cycles. The formal correspondence to ordinary QM obtained by imposing intrinsic periodicity to elementary particles strongly suggests that relativistic QM can be interpreted as an effective (statistical) description of semi-classical mechanics associated to the fast cyclic dynamics of elementary particles, as also suggested for instance by 't Hooft --- and not the contrary. We have proven that fundamental aspects of QM can be consistently described in terms of the harmonic modes of four-dimensional vibrating `strings'. The resulting theory is a relativistic generalization of the theory of sound (where sources can vibrate along the time dimension). This aspect anticipates the study of QM  to Pythagoras, who first investigated the ``quantized'' spectrum of harmonic system, as also recently suggested by Wilczek. The resulting theory conciliates together important aspects of modern physics: it can be regarded as a (purely 4D) string theory (defined on the single compact word-parameter $s$) similarly to original Veneziano's proposal; it provides a geometrodynamical description of gauge interaction, similarly to  original Weyl's proposal; it pinpoints fundamental aspects (e.g. classical to quantum correspondence) of Maldacena's duality, originally pointed out by Witten; it justifies the mathematical beauty of extra-dimensional theories as original proposed by N\"ordstrom, Kaluza and Klein; and so on. We conclude that physical origin (the ``missing link'' \cite{Ferber :1996}) of QM must be found in the assumption of intrinsic periodicity, as originally proposed by the father's of QM and then forgotten. The resulting formulation of physics in terms of elementary space-time cycles brings important new elements to face the open problems of modern physics \cite{Dolce:2009ce,Dolce:tune,Dolce:AdSCFT}.

\section*{References}
\bibliographystyle{iopart-num}

\begin{thebibliography}{1}
\expandafter\ifx\csname url\endcsname\relax
  \def\url#1{{\tt #1}}\fi
\expandafter\ifx\csname urlprefix\endcsname\relax\def\urlprefix{URL }\fi
\providecommand{\eprint}[2][]{\url{#2}}

\bibitem{Dolce:2009ce}
Dolce D 2011 {\em Found. Phys.\/} {\bf 41} 178 (\textit{Preprint}
  \eprint{0903.3680v5})

\bibitem{Dolce:tune}
Dolce D 2012 {\em Ann. Phys.\/} {\bf 327} 1562 (\textit{Preprint}
  \eprint{1110.0315}) 

\bibitem{Dolce:AdSCFT}
Dolce D 2012 {\em Ann. Phys.\/} {\bf 327} 2354 (\textit{Preprint}
  \eprint{1110.0316}) 

\bibitem{Broglie:1924}
Broglie L~d 1924 {\em Phil. Mag.\/} {\bf 47} 446

\bibitem{'tHooft:2001ar}
't~Hooft G 2003 {\em Int. J. Theor. Phys.\/} {\bf 42} 355
  (\textit{Preprint} \eprint{hep-th/0104080})

\bibitem{Elze:2003ws}
Elze H~T 2004 {\em Lect. Notes Phys.\/} {\bf 633} 196 (\textit{Preprint}
  \eprint{gr-qc/0307014})

\bibitem{2008FoPh...38..659C}
{Catillon} P, et.al. 2008 {\em
  Found. Phys.\/} {\bf 38} 659

\bibitem{Jaffe:2005vp}
Jaffe R~L 2005 {\em Phys. Rev. D\/} {\bf 72} 021301 (\textit{Preprint}
  \eprint{hep-th/0503158})

\bibitem{Solovev:2011}
Solov'ev E 2011 {\em Eur. phys. J. D\/} {\bf 65} 331--351

\bibitem{Ferber :1996}
Ferber R. 1996 {\em Found. Phys. Lett.\/} {\bf 9} 6, 575


\end{thebibliography}
\providecommand{\newblock}{}

\end{document}